\documentclass[nohyper, letterpaper,11pt,notoc]{JHEP3}

\usepackage{amsfonts,amsmath,amssymb}
\usepackage{bm,bbm}
\usepackage{graphicx,epsfig}

\newcommand{\gsim}{\mathrel{\lower0.8ex\vbox{\lineskip=0.15ex\baselineskip=0ex
                   \hbox{$>$}\hbox{$\sim$}}}}
\newcommand{\lsim}{\mathrel{\lower0.8ex\vbox{\lineskip=0.15ex\baselineskip=0ex
                   \hbox{$<$}\hbox{$\sim$}}}}

\newcommand{\sla}[1]{{\raise.15ex\hbox{$/$}\kern-.57em #1}}
\newcommand{\Sla}[1]{\kern0.12em{\raise.15ex\hbox{$/$}\kern-.74em #1}}
\newcommand{\del}{\partial}
\newcommand{\tr}{{\rm tr}}
\newcommand{\wbar}[1]{\overline{#1}}
\newcommand{\wtild}[1]{\widetilde{#1}}
\newcommand{\dubdel}{\!\!\stackrel{\leftrightarrow}{\raise.02ex\hbox{$\del$}}}
\newcommand{\dubD}{\!\!\stackrel{\leftrightarrow}{\raise.02ex\hbox{$D$}}}
\newcommand{\ddel}{\!\mathop{\vbox{\lineskip=0ex\baselineskip=0ex
                   \hbox{{\small $\leftrightarrow$}}
                   \hbox{$\hspace{.2ex}\partial$}}}\hspace{-.96ex}}

\newcommand{\DD}{\!\mathop{\vbox{\lineskip=0ex\baselineskip=0ex
                   \hbox{{\small $\leftrightarrow$}}
                   \hbox{$\hspace{.0ex}D$}}}\hspace{-.53ex}}

\newcommand{\kev}{\>{\rm keV}}
\newcommand{\Mev}{\>{\rm MeV}}
\newcommand{\Gev}{\>{\rm GeV}}

\newcommand{\pb}{\>{\rm pb}}

\newcommand{\beq}{\begin{eqnarray}}
\newcommand{\eeq}{\end{eqnarray}}
\newcommand{\nn}{\nonumber}
\newcommand{\eql}[1]{\label{eq:#1}}
\newcommand{\eq}[1]{(\ref{eq:#1})}

\newcommand{\varep}{\varepsilon}

\newcommand{\SU}{{\rm SU}}
\newcommand{\U}{{\rm U}}

\newcommand{\col}{{\tilde{\rho}}}
\newcommand{\ourpi}{{\tilde{\pi}}}

\preprint{UMD-PP-08-002}

\title{Colored Resonances at the Tevatron:\linebreak Phenomenology and Discovery
Potential in Multijets}

\author{Can Kilic$^{(a)}$, Takemichi Okui$^{(a,b)}$ and Raman Sundrum$^{(a)}$
\\ \it{(a) Department of Physics and Astronomy, Johns Hopkins University,\\Baltimore, MD 21218, USA}
\\ \it{(b) Department of Physics, University of Maryland,\\College Park, MD 20742, USA}}

\abstract{There exist several classes of theories beyond the Standard Model which contain massive spin-1 color octets, generically called ``colorons''.
Indeed we argue that colorons inevitably appear in the spectrum whenever new colored particles feel an additional confining force. Colorons are
distinctive at hadron colliders as this is the only environment in which they can be resonantly produced. In the simplest models we show that the coloron
naturally decays to multijets via secondary resonances, which can be consistent with all existing bounds, even for colorons as light as a few hundred GeV.
We perform representative case studies and show that a search in the four-jet channel at the Tevatron has strong signal significance, while the LHC faces
formidable challenges for such a search.}


\begin{document}
\section{Introduction and the Scenario}
\label{sec:intro}

Hadron colliders such as the Tevatron and LHC are ideal for producing new colored particles with weak-scale masses. Such particles may be associated with
mechanisms for resolving the Hierarchy Problem, such as Supersymmetry, Strong Dynamics, Extra Dimensions, or Little Higgs, or they may arise for other
seemingly incidental reasons (as did the second and third generation quarks).  The standard strategy for discovery of such states is to take advantage of
QCD cross sections for their {\it pair-production}, and to exploit their electroweak or other distinctive interactions in order to pull them out of the
enormous QCD backgrounds. New electroweak and/or flavor physics is however strongly constrained by a host of precision data, which hints that such new
states are relatively heavy. A different possibility arises if there is a new colored particle with the same quantum numbers as the gluon, a color-octet
massive vector boson, which can then be produced {\it resonantly} via mixing with a virtual $s$-channel gluon.  One can then hope to pick out such a
sizable resonance from the non-resonant QCD backgrounds, even without relying on flavor-tagged, leptonic, or missing energy signatures.

Indeed color octet vector particles have appeared in new physics proposals in various guises: massive gauge bosons from extensions of QCD gauge structure
in Topcolor models \cite{coloron-1}, Kaluza-Klein excited gluons in extra-dimensional models, composite colored vector mesons in non-minimal Technicolor
\cite{Farhi:1979zx}, and string excitations of the gluon in TeV Gravity \cite{TeV-string}. We will borrow the terminology of Refs.~\cite{coloron-1,
coloron-2} and refer to any such massive vector particle as a ``coloron''.

Colorons can also arise rather minimally, apart from the dramatic scenarios cited above. Imagine QCD pair-production of a {\it new} colored particle and
its anti-particle via $s$-channel gluon exchange, but where the pair are bound together with a new stronger force.  The bound state is then necessarily a
coloron! This is highly plausible given the prevalence of confinement among non-abelian gauge theories.  In fact, Nature has given us a useful precedent
in QED and hadronic physics. A virtual photon from electron-positron annihilation can pair-produce new charged particles and anti-particles. When the new
particles happen to be quarks, the strong interactions can confine the quark and anti-quark pair into a single resonance, the $\rho$, emerging from the
virtual photon. In hadronic physics this is referred to as ``photon-$\rho$ mixing''.%
\footnote{To be clear then, in this analogy the role of the virtual gluon
created by a $q$-$\bar{q}$ pair is played by a virtual photon created by an
$e^+$-$e^-$ pair, the role of the composite coloron is played by the composite $
\rho$ meson, and the role of a new weak-scale strong force holding the
coloron together is played by the ordinary GeV-scale strong interactions of QCD
holding the $\rho$ together.}

Our viewpoint is that the coloron is an object of general phenomenological interest, much like a $Z'$, readily produced and with diverse theoretical
motivations, and that  search strategies should be devised to cover the promising signatures. In this paper, we point out that there is a sizable regime,
which is not excluded by existing data, where a coloron can be discovered in multi-jet studies at the Tevatron, which is however \textit{much more
difficult to find at the LHC}. Discovery does not require the new physics to carry electroweak or flavor quantum numbers, and therefore the Tevatron
accessibility is not necessarily in conflict with precision data. In this paper this is naturally achieved by having the quarks couple to the coloron only
via gluon-coloron mixing. Stronger flavor blind quark-coloron couplings have been proposed earlier in \cite{coloron-2}, however this scenario is excluded
in the sub-TeV regime which is our focus \cite{Simmons:1996fz,Choudhury:2007ux}. While precision constraints do allow the coloron to couple strongly to
the top quark \cite{Hill:1993hs,Dicus:1994sw} ($t_{R}$ in particular), at Tevatron energies this is severely constrained by measurements of the top
production cross section. Ref.~\cite{Casalbuoni:1995rg} studied the early Tevatron implications of a phenomenological model containing a coloron with an
adjustable coupling to the top quark. For the LHC phenomenology of heavier colorons with strong top coupling, see
Refs.~\cite{LHC-colorons,Lillie:2007hd,Guchait:2007jd}.

In fact the most severe constraints come, not from electroweak or flavor tests, but from past dijet studies \cite{dijet-UA2, dijet-Teva, bb-dijet}.  A
coloron, once produced, can decay by mixing back into a virtual gluon and then into $q$-$\bar{q}$, thereby creating a dijet resonance.  If this channel
dominated the decay of the coloron, then most of the sub-TeV mass range is already excluded (for a ``typical'' size of gluon-coloron mixing, to be
discussed shortly). However, looking at it more closely this is not what the QED analogy with the $\rho$ meson suggests.  Notice that $\rho \to e^+ e^-$
is {\it not} the dominant decay mode of $\rho$ at all; it is $\rho \to \pi\pi$ that is nearly 100\%.  Translated to the case of interest, the coloron can
naturally decay dominantly into other new colored resonances, which may in turn decay into several jets, thereby diminishing the dijet decays of the
coloron below existing bounds.  The coloron effectively becomes a multi-jet resonance. The purpose of this paper is to study the phenomenology of the
coloron and secondary resonances in multi-jet processes.

The paper is organized as follows: In section \ref{sec:model} we write down a simple and renormalizable theory of a coloron and a secondary resonance
realized as bound states of new strong dynamics in close analogy to hadronic physics. In section \ref{sec:eff-Lag} we introduce a phenomenological
Lagrangian for this theory capturing the production of the coloron as well as subsequent decays. We will then show in section \ref{sec:const} that this
benchmark model is consistent with existing constraints and in section \ref{sec:discov} will lay out a search strategy for its discovery at the Tevatron
for a range of coloron masses. We will conclude in section \ref{sec:conc} with a discussion of our results and future prospects for alternative coloron
models.

\section{An Illustrative Model}
\label{sec:model}

The purpose of the following simple field theory model is to demonstrate how general the phenomenon of a coloron can be, that it can be decoupled from
dangerous new electroweak or flavor effects, and to show that the coloron can readily have strong decay modes that dilute the dijet resonance channel,
making it predominantly a multi-jet signal.

Consider the renormalizable theory given by
\beq
  {\cal L}
  = {\cal L}_{\rm SM} + \bar{\psi} (i \Sla{D} - m) \psi
   -\frac{1}{4} H_{\mu \nu} H^{\mu\nu}  \>.
\eql{UV-Lag}
\eeq
The second term simply describes the addition of some new color-triplet Dirac fermions, such as we might contemplate pair-producing at a hadron collider.
For maximal safety in precision tests let us take them to be electroweak singlets with no connection to standard-model (SM) flavor. In addition we assume
that the new fermions transform under a new non-abelian gauge group, ``hyper-color'' (HC), described by the third term above, which is confining at a
scale $\Lambda_{\rm HC}$. New confining forces, in particular modeled on QCD, play a role in a number of other phenomenologically interesting scenarios
beyond the SM, such as Technicolor \cite{techni} and its variants, ``hidden valley'' \cite{hidden}, or ``quirk'' models \cite{Kang:2008ea}.

Our analysis will be relevant when $\Lambda_{\rm HC}$ is in the range of $100$s of GeV, but it is not fundamentally constrained to that regime.
Therefore, instead of pair-production of the new hyper-quarks we will have production of the $\Lambda_{\rm HC}$-scale hyper-hadrons.  We can in fact
safely set the ``current'' mass to zero, $m \rightarrow 0$, and we do so from now on for simplicity.

The fact that the new particles are all flavor and electroweak ``blind'' ensures
their safety for most precision tests.  (There is a blind precision test
however, the tests of quark ``compositeness'' effects, that will be treated
separately in section \ref{sec:const}.)  The real meaning behind the technical
renormalizability of the model is  that there can be a separation of scales
between $\Lambda_{\rm HC}$ and yet other new physics.  Such new physics may have
electroweak and flavor components, and may address the hierarchy problem, but it
does not have to interfere with the lower-energy physics of the HC
sector.

For convenience, we take HC to mimic QCD structure as much as possible so
that we can translate strong-interaction matrix elements directly from
their measured QCD values to the HC physics, with the simple rescaling
$\Lambda_{\rm QCD} \rightarrow \Lambda_{\rm HC}$.  We therefore take the
HC gauge group to be $\SU(3)_{\rm HC}$, with three massless flavors, $\psi$,
and therefore $\SU(3)$ flavor symmetry. This is just a rescaled version of
QCD with three light flavors, without current masses or electroweak
charges. An important difference  is that
while QCD's flavor symmetry is weakly gauged, but only partially, by
electromagnetism, HC's flavor symmetry will be completely gauged by
QCD itself (which is weak at $\Lambda_{\rm HC}$ energies).  That is,
we are identifying HC flavor symmetry with QCD gauge symmetry, thus
taking our Dirac fermions, $\psi$, to be bi-fundamentals
$({\bf 3}, {\bf \bar{3}})$ of $\SU(3)_{\rm QCD} \times \SU(3)_{\rm HC}$.

We can therefore automatically estimate the spectrum of the HC sector by just rescaling the ordinary hadronic spectrum. In particular, we know that there
is an $\SU(3)$ flavor octet of massive vector mesons imitating the ordinary $ \rho$ (and its flavor siblings).  Since the $\SU(3)$ flavor symmetry of the
hyper-hadrons is identified with $\SU(3)_{\rm QCD}$, this is precisely a QCD octet ``coloron'', which is created by the QCD color current (virtual gluon)
just as the $\rho^0$ is created by the electromagnetic current (virtual photon). The coloron mass is of order $\Lambda_{\rm HC}$, and we can just use
$m_{\rm coloron}$ as our unit of measure rather than $\Lambda_{\rm HC}$. The coloron is not the lightest hyper-hadron however, there must be a QCD octet
hyper-pion, which is lighter than the coloron, being a (pseudo-)Goldstone multiplet of the HC dynamics, as we will see more explicitly below. In
particular the coloron can decay into a hyper-pion pair. While there will be a host of heavier hyper-hadrons they will not be relevant for our analysis or
Tevatron studies. They may however be interesting for the LHC.

Non-minimal technicolor models \cite{Farhi:1979zx} also contain a coloron with a colored techni-pion decay mode. Ref.~\cite{Eichten:1984eu} considered the
SSC phenomenology of techni-pion production but found the coloron resonance relatively uninteresting. Ref.~\cite{Lane:2002sm} devoted to walking
technicolor phenomenology took the couplings for coloron decay to techni-pions to be insignificant. Ref.~\cite{Zerwekh:2001uq} studied coloron production
and techni-pion decays at the Tevatron, but focussed on a signal with a photon in the final state, which was found not to be significant enough for
discovery.

\section{The Phenomenological Lagrangian}
\label{sec:eff-Lag}

At the kind of precision and energy at the Tevatron, we can approximate the
complexity of the full HC dynamics by simple effective vertices for the
coloron, $\col$, and hyper-pion, $\ourpi$, as well as their couplings to QCD,
sufficient to describe their production and decays.  We can then fit the
couplings of the effective vertices by matching them to the analogous vertices
describing the ordinary $\rho$ and $\pi$ mesons and their couplings to QED,
which can be extracted from hadronic data.

Omitting the electroweak-Higgs sector and the leptons, our effective
Lagrangian is given by
\beq
  {\cal L}_{\rm eff}^{\rm HC}
  &=& \wbar{q} i\Sla{\wtild{D}} q
     -\frac14 G^a_{\mu\nu} G^{a\mu\nu}  \nn\\
  && -\frac14 \col^a_{\mu\nu} \col^{a\mu\nu}
     +\frac{m_\col^2}{2} \col^a_\mu \col^{a\mu}
     +\frac{\tilde{\varep}}{2} \col^a_{\mu\nu} G^{a\mu\nu}  \nn\\
  && +\frac12 (\wtild{D}_\mu \ourpi)^a (\wtild{D}^\mu \ourpi)^a
     -\frac{m_\ourpi^2}{2} \ourpi^a \ourpi^a  \eql{HC-eff}\\
  && -g_{\col\ourpi\ourpi} f^{abc} \col^a_\mu \ourpi^b \wtild{D}^\mu \ourpi^c
     -\frac{3g_3^2 \epsilon^{\mu\nu\rho\sigma}}{16\pi^2 f_\ourpi}
      \tr\bigl[ \ourpi^{a}T^{a} G_{\mu\nu} G_{\rho\sigma} \bigr]  \>.\nn
\eeq
Here, the indices $a$, $b$, $c$ run from 1 through 8, labelling the eight states of the color octet.  $\wtild{D}_\mu$ denotes the covariant derivative
containing a gluon field, and the reason for the tilde will become clear shortly.  The first line above describes the kinetic terms for the standard-model
quark $q$ and the gluon $G_{\mu\nu} = G^a_{\mu\nu} T^a$, with the $\SU(3)$ generators $T^a$.  The second line contains the kinetic and mass terms for the
coloron, where $\col_{\mu\nu} \equiv \wtild{D}_\mu\col_\nu -\wtild{D}_\nu\col_\mu$, and the gluon-coloron mixing parameterized by $\tilde{\varep}$. The
third line describes the kinetic and mass terms for the hyper-pions.  The fourth line describes the $\col \to \ourpi\ourpi$ decay and the $\ourpi \to gg$
decay.

Now, note that the above phenomenological Lagrangian has a close analog in
Nature:
\beq
  {\cal L}_{\rm eff}^{\rm QCD}
  &=& \wbar{e} i\Sla{D} e -\frac14 F_{\mu\nu} F^{\mu\nu}  \nn\\
  && -\frac14 \rho_{\mu\nu} \rho^{\mu\nu}
     +\frac{m_\rho^2}{2} \rho_\mu \rho^{\mu}
     +\frac{\varep}{2} \rho_{\mu\nu} F^{\mu\nu}  \nn\\
  && +\frac12 \del_\mu \pi^0 \del^\mu \pi^0 - m_{\pi^0}^2 \pi^0 \pi^0  \nn\\
  && +D_\mu \pi^- D^\mu \pi^+ - m_{\pi^{\pm}}^2 \pi^- \pi^+  \eql{QCD-eff}\\
  && -ig_{\rho\pi\pi} \rho^\mu (\pi^- \DD_\mu \pi^+)
     -\frac{e^2 \epsilon^{\mu\nu\rho\sigma}}{32\pi^2 f_\pi}
      \pi^0 F_{\mu\nu} F_{\rho\sigma}  \nn\>,
\eeq
where $A \ddel B \equiv A\,\del B - (\del A) B$. Here, $D_\mu $ denotes the covariant derivative containing a photon, and the absence of the tilde
distinguishes it from $\wtild{D}_\mu$. The first line describes the kinetic terms for the electron and the photon, while the second line contains the
kinetic and mass terms for the neutral $\rho$ meson, where $\rho_{\mu\nu} \equiv \del_\mu\rho_\nu -\del_\nu\rho_\mu$, and the photon-$\rho$ mixing
parameterized by $\varep$.  The third and fourth lines describe the kinetic and mass terms for the neutral and charged pions, and finally the fifth line
describes the $\rho \to \pi^+\pi^-$ decay and the $\pi^0 \to \gamma\gamma$ decay.  (Note that $f_\ourpi$ and $f_\pi$ are normalized in the same way.)

Now, let us translate carefully between \eq{HC-eff} and \eq{QCD-eff} to extract the parameters in \eq{HC-eff} from \eq{QCD-eff}.  First, there is a
straightforward change of the scale, from $\Lambda_{\rm QCD} \sim m_\rho$ to $\Lambda_ {\rm HC} \sim m_\col$.  This immediately implies that $f_\pi \simeq
92\Mev$ is translated to
\beq
  f_\ourpi \simeq 92\Gev \frac{m_\col}{10^3 m_\rho}  \>,
\eql{f_pi}
\eeq
although the precise value of $f_\ourpi$ is not important for our phenomenology,
since all we need to know is that $\ourpi$ decays promptly.

We now turn to $g_{\col\ourpi\ourpi}$. Since QCD is a small perturbation to the hyper-color dynamics just like QED is a small perturbation to the color
dynamics, we expect that $g_{\col\ourpi\ourpi} = g_{\rho\pi\pi}$ as the decay $\rho \to \pi^+\pi^-$ is governed by the strong interactions.  All we need
to check is the normalization conventions, for which considering the entire octet of the mesons is useful:
\beq
    f^{abc} \rho^a_\mu \pi^b \del^\mu \pi^c
  = i\rho^\mu (\pi^- \ddel_{\mu} \pi^+)
    + \cdots  \>,
\eeq
which shows $g_{\col\ourpi\ourpi}$ and $g_{\rho\pi\pi}$ are normalized in the
same way.  Thus, we have
\beq
  g_{\col\ourpi\ourpi} = g_{\rho\pi\pi} \simeq 6  \>,
\eql{g_rhopipi} \eeq
which is extracted by using \eq{QCD-eff} to fit $\Gamma_{\rho \to \pi\pi} = 149
\Mev$. As we will show in the next section, this strong coupling puts us
significantly below any dijet bounds.

Next, to make the $\col$-$q$-$\bar{q}$ coupling explicit, we redefine the gluon field as $G^a_\mu \to G^a_\mu + \tilde{\varep} \col^a_\mu$. Neglecting
$O(\tilde{\varep}^2)$, this eliminates the $\tilde{\varep}$ term in the second line of \eq{HC-eff}.  However, the shift also affects $\wbar{q}
i\Sla{\wtild{D}} q$, and the following new term now appears in \eq{HC-eff}:
\beq
  {\cal L}_{\col q \bar{q}}
  = -g_3 \tilde{\varep} \, \col^a_\mu \, \wbar{q} \gamma^\mu T^a q
    \>,
\eql{rho-q-q}
\eeq
which describes the coloron production from a $q$-$\bar{q}$ collision and its decay into a dijet.  To determine the value of $\tilde{\varep}$, we perform
the analogous shift in \eq{QCD-eff}, which induces the $\rho$-$e^+$-$e^-$ coupling $-e\varep \rho_\mu \wbar{e} \gamma^\mu e$.  Using this to fit the
partial width $\Gamma_{\rho \to e^+e^-} = 7.04\kev$ gives $\varep \simeq 0.06$.  To translate this to $\tilde{\varep}$, notice that the $\rho$-$\gamma$
mixing parameter $\varep$ itself is proportional to the gauge coupling $e$, because ``microscopically'' it is the quark-antiquark pair inside the $\rho$
annihilating into the (off-shell) photon.  Thus, the $\tilde {\varep}$ must be rescaled by the ratio of the gauge couplings:
\beq
  \tilde{\varep} = \frac{g_3}{e} \varep \simeq 0.2  \>,
\eql{mixing}
\eeq

The last parameter to be extracted is $m_\ourpi$.  Note that the hyper-pions $ \ourpi$ would be exact Goldstone bosons if $g_3$ were zero.  The
corresponding statement in the analog \eq{QCD-eff} is that the mass-squared {\it difference} between the charged and the neutral pions would be zero if
$e$ were zero (up to small corrections of $O(m_{u,d}^2)$). Therefore we can use chiral perturbation theory to extrapolate the mass of the hyper-pion from
the pion mass difference. We find
\beq
  \frac{m_{\ourpi}^2}{m_\col^2} = 3\,\frac{g^2_3}{e^2}\,\frac{m_{\pi^\pm}^2-m_{\pi^0}^2}{m_\rho^2}
\eeq
where we have included the color factor. Numerically this gives
\beq
  m_\ourpi \simeq 0.3 m_\col  \>.
\eql{pi-mass}
\eeq

To summarize, we will in the rest of the paper use the effective Lagrangian
\beq
  {\cal L}_{\rm eff}^{\rm HC}
  &=& \wbar{q} i\Sla{\wtild{D}} q
     -g_3 \tilde{\varep} \, \rho^a_\mu \, \wbar{q} \gamma^\mu T^a q
      \nn\\
  && -\frac14 G^a_{\mu\nu} G^{a\mu\nu}
     -\frac14 \col^a_{\mu\nu} \col^{a\mu\nu}
     +\frac{m_\col^2}{2} \col^a_\mu \col^{a\mu}  \nn\\
  && +\frac12 (\wtild{D}_\mu \ourpi)^a (\wtild{D}^\mu \ourpi)^a
     -\frac{m_\ourpi^2}{2} \ourpi^a \ourpi^a  \eql{eff-Lag}\\
  && -g_{\col\ourpi\ourpi} f^{abc} \col^a_\mu \ourpi^b \del^\mu \ourpi^c
     -\frac{3g_3^2 \epsilon^{\mu\nu\rho\sigma}}{16\pi^2 f_\ourpi}
      \tr\bigl[ \ourpi^{a}T^{a} G_{\mu\nu} G_{\rho\sigma} \bigr]  \>,\nn
\eeq
where $f_\ourpi$, $g_{\col\ourpi\ourpi}$, $\tilde{\varep}$, and $m_\ourpi$ are given by \eq{f_pi}, \eq{g_rhopipi}, \eq{mixing}, and \eq{pi-mass}, this
will be referred to as ``the benchmark model''. Note that in the benchmark model $m_\col$ is the only free parameter. In this paper we will restrict
ourselves to a range for $m_{\col}$ which makes the coloron discoverable at the Tevatron. We will elaborate further on this in our conclusions. A case
study for the discovery potential will be presented in section \ref{sec:discov} with a strong result.

We should mention at this point that while it is not possible to resonantly produce a coloron from a gluon-gluon initial state through renormalizable
operators, there are higher dimensional operators which can do this, the leading one being
$(\alpha_{s}/m_{\col}^{2})f^{abc}\col^{a\,\mu}_{\nu}G^{b\,\sigma}_{\mu}G^{c\,\nu}_{\sigma}$, using naive dimensional analysis
\cite{Manohar:1983md,Georgi:1985kw} for the strong hypercolor interactions. However, the effect of this operator on resonant coloron production is
negligible at the Tevatron since the coupling to $q\bar{q}$ combined with valence quark PDF's completely dominates the cross section (so that a more
precise estimate of the coefficient of the above operator is unnecessary).

\FIGURE[!t]{\includegraphics[width=5.0in]{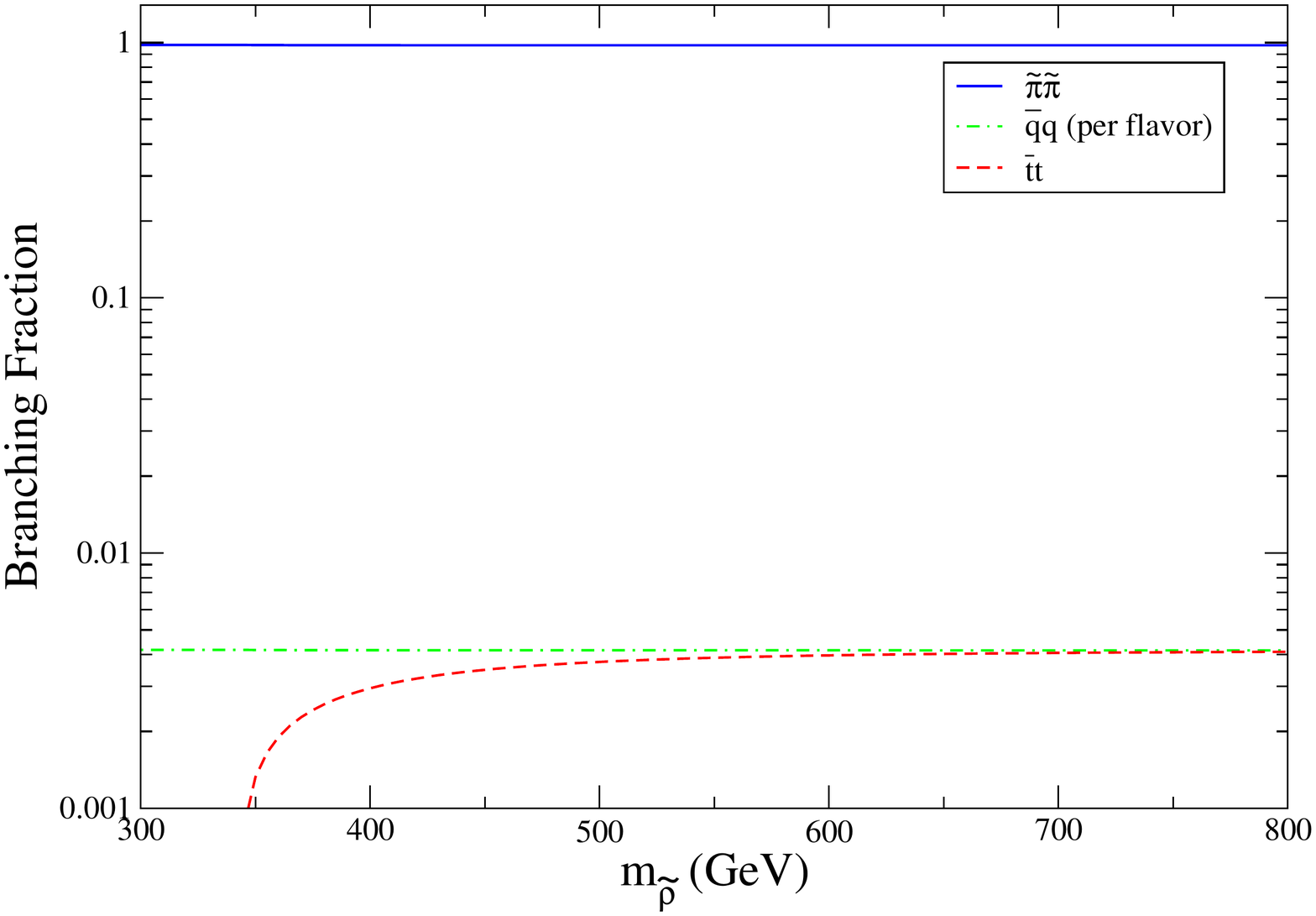} \caption{The branching fractions of the coloron as a function of its mass in the benchmark model.}
\label{fig:bf}}
As we have alluded to in the introduction, in the benchmark model the dominant decay mode of the coloron is into a pair of hyper-pions, and the branching
fraction into quarks is suppressed by the mixing of the coloron with the gluon, which is the reason why this model is not in conflict with the dijet
resonance bounds from the Tevatron. This will be explained in detail in section \ref{sec:const}. In figure \ref{fig:bf} we plot the branching fractions of
the coloron as a function of its mass, calculated using COMPHEP4.4 \cite{MC}.

It should be kept in mind that while the parameters appearing in Eq.~\eq{eff-Lag} are ultimately determined by the theory in Eq.~\eq{UV-Lag}, one can
easily imagine that alternate fundamental dynamics can lead to different values for these parameters. Physically $\tilde{\varep}$ sets the overall
production cross section of the coloron, while the ratio $\tilde{\varep}/g_{\col\ourpi\ourpi}$ sets the ratio of the partial decay widths $\Gamma_{\col\to
q\bar{q}}/\Gamma_{\col\to\ourpi\ourpi}$. In section \ref{sec:const} we will quantitatively relate the constraints on the benchmark model to these
effective parameters, as well as $m_{\ourpi}$.

\section{Constraints on the Benchmark Model}
\label{sec:const}

In this section we will go through various potential constraints on the
benchmark model described in the previous section, and establish that our
scenario
is compatible with existing experimental bounds. The benchmark model, with $g_
{\col\ourpi\ourpi}$ and $m_{\ourpi}/m_{\col}$ fixed, has only one free
parameter, namely $\Lambda_{\rm HC}$ or equivalently $m_\col$.

\subsection{Constraints on the $\col$ Particle}
\label{sec:rho-const}

\FIGURE[!t]{\includegraphics[width=5.0in]{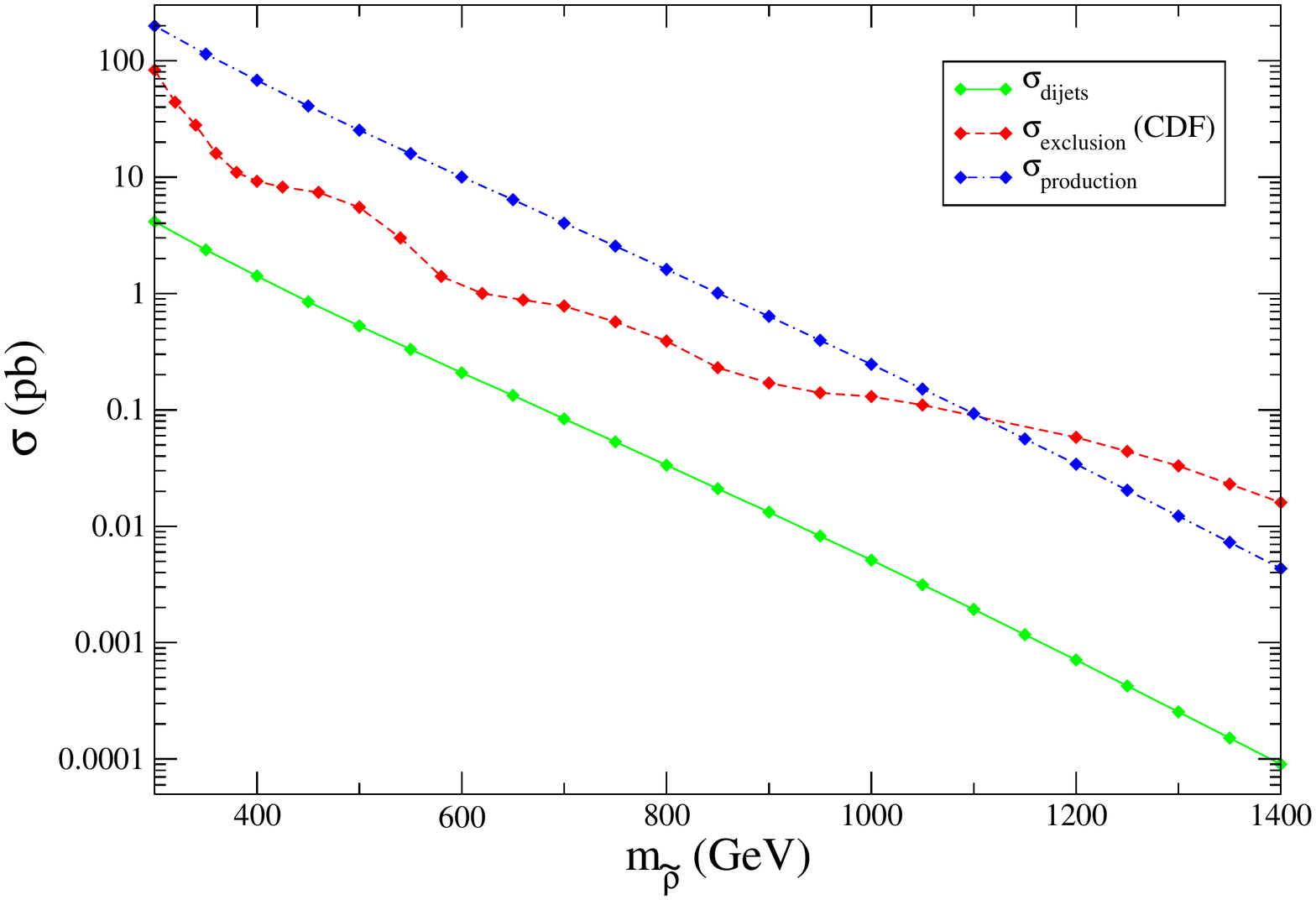} \caption{The comparison of our benchmark model with recent CDF exclusion bounds on dijet
resonances. The green curve represents the cross section for dijet production through the coloron at Tevatron Run-II as a function of the coloron mass and
the red curve represents the CDF dijet exclusion bound obtained from $1.13~\rm{fb}^{-1}$ of data. In this plot signal is presented with perfect acceptance
while the exclusion curve was obtained by demanding that both jets be central ($|y^{\rm{jet}1,2}|<1$), thus in reality our benchmark model is even less
constrained than this conservative plot suggests. For completeness we include the \textit{total} production cross section of the coloron represented by
the blue curve which illustrates that for the coloron not to be excluded, the smallness of the dijet branching fraction is crucial.}
\label{fig:dijetexclusion}}
An obvious constraint on the coloron comes from resonance searches in the dijet channel. The most recent publicly available bounds on resonant dijet
production are reported in \cite{dijet-Teva} as well as \cite{bb-dijet} (for heavy flavor-tagged jets). We plot in figure \ref{fig:dijetexclusion} the
dijet production cross section through the coloron (calculated using \cite{MC}) in the benchmark model as a function of $m_\col$ and compare to the bounds
obtained by the CDF collaboration. We remark here that the exclusion curve has detector acceptance folded in (both jets are required to be central,
$|y^{\rm{jet}1,2}|<1$) while for the signal we are plotting the cross section assuming perfect acceptance, hence this plot is overly conservative. We
include in this plot the \textit{total} $\col$ production cross section, note that for most choices of $m_{\col}$ the model would have been excluded if
dijets were the dominant decay mode of the coloron. However, the presence of the $\ourpi$-$\ourpi$ mode lowers the dijet production cross section
significantly below the bound. A potential worry, namely that detector effects may cause a fraction of the four-jet signal events to be reconstructed as
dijet events with the correct value of $m_{\col}$, is disarmed by the fact that the total production cross section of the coloron is always within a
factor of order one of the dijet bound.

The dijet bound is also the primary constraint for going beyond the benchmark model and varying the parameters of the phenomenological Lagrangian
presented in \eq{eff-Lag}. It is straightforward to verify that for a given value of $m_{\col}$ the dependence of the dijet cross section on the model
parameters is given by
\beq \sigma_{dijet}=\left(\sigma_{0}\right)_{dijet}\left(\frac{\tilde{\varep}}{\tilde{\varep}_{0}}\right)^{4}
\left(\frac{g_{\col\ourpi\ourpi,0}}{g_{\col\ourpi\ourpi}}\right)^{2}\left(\frac{m_{\col}^{2}-4m_{\ourpi,0}^{2}}{m_{\col}^{2}-4m_{\ourpi}^{2}}\right)^{3/2}
\eeq
(assuming $\Gamma_{\col\to q\bar{q}}\ll \Gamma_{\col\to\ourpi\ourpi}$ still holds) where the subscripts 0 denote the parameters of the benchmark model.
(For $\Gamma_{\col\to q\bar{q}}\gg \Gamma_{\col\to\ourpi\ourpi}$, $\sigma_{dijet}\sim\sigma_{prod}$.) Thus, it is straightforward to use figure
\ref{fig:dijetexclusion} to constrain the parameters of \eq{eff-Lag}.

\FIGURE[!t]{\includegraphics[width=5.0in]{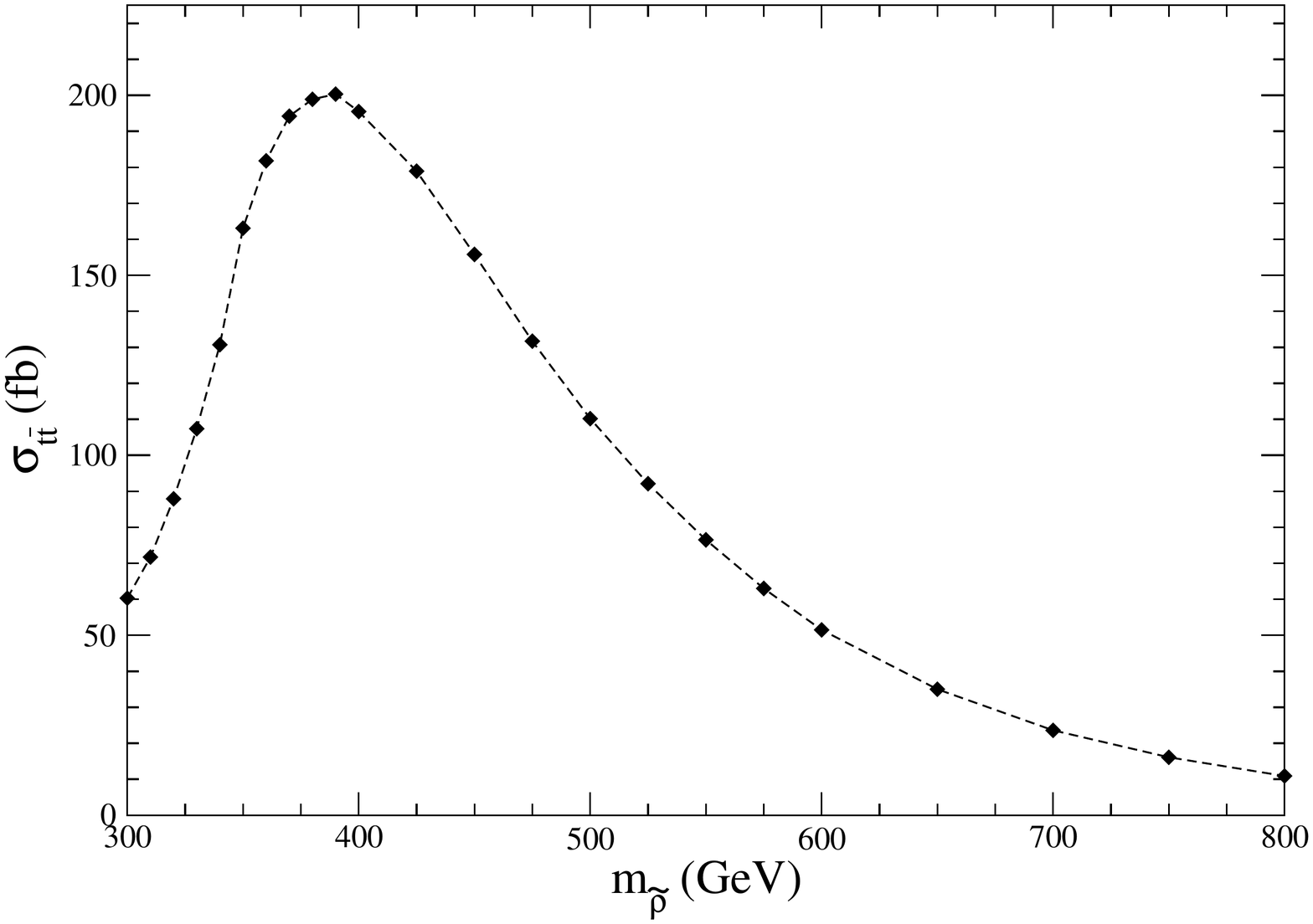} \caption{$\sigma (p\bar{p} \to \col \to t\bar{t})$ cross section at Tevatron Run-II as a
function of $m_\col$ for the benchmark model. For $m_{\col}$ near or below the $t\bar{t}$ threshold, we plot the cross section with one of the $t $-quarks
off-shell.} \label{fig:ttbar}}
The $t$-$\bar{t}$ branching mode is another source of potential constraints on the $\col$ production cross section. In figure \ref{fig:ttbar} we plot the
$t$-$\bar{t}$ cross section via $\col$ production and decay as a function of $m_ \col$ in the benchmark model (calculated using \cite{MC}). Note that the
cross section stays below $0.2\pb$ for the entire range of $m_{\col}$, which is below the lower bounds in \cite{ttbar}.  Note that \cite{ttbar} searches
for a {\it narrow} resonance that decays to a $t$-$\bar{t}$ pair, so the bound on a wide resonance such as our $\col$ is actually even weaker.

There are studies of multi-jet final states at Tevatron Run-I \cite{multi-jets}
which found no deviations from the QCD predictions, however these studies
use large $p_{T}$ and $m_{\rm inv}$ cuts such that the events coming from
a light coloron ($m_{\col}<500\Gev$) do not pass the analysis cuts while for a
heavier coloron the cross section is low enough such that any excess produced is
not statistically significant. We have found $m_{\col}\sim700\Gev$ to be
the point where the number of events passing the cuts used in \cite{multi-jets}
is maximized at roughly 60, which would correspond to a $2\sigma$ excess
in their distributions.

In our benchmark model, there is a three jet decay mode $\col\rightarrow g\ourpi$ for the coloron, which is the analogue of $\rho\rightarrow\gamma\pi$ in
QCD. Scaling up the partial width for this process from QCD using the appropriate factors ($\frac{\alpha_{s}}{\alpha_{em}}$ for coupling constants and 3
for number of colors) we find that the relevant branching fraction is a few percent, so the number of three jet events from this decay mode should be
similar to the number of dijet events. Ref.~\cite{multi-jets} is insensitive to such a low number of three jet events.

Finally, VISTA and SLEUTH global searches \cite{Vista-Sleuth} have been performed to look for anomalies in the Tevatron data (with emphasis on high-$p_T$
deviations). As we will show in section \ref{sec:discov} a blind global search has limited sensitivity to the presence of $\col$ while a more optimized
search taking advantage of the presence of secondary resonances yields much stronger evidence for a discrepancy in kinematic distributions.

\subsection{Constraints on the $\ourpi$ Particle}
\label{sec:resonant-pi}

The $\ourpi$ particle in the benchmark model has a coupling to a pair of gluons through the anomaly, and can thus be resonantly produced from a $g$-$g$
initial state. The $\ourpi$ subsequently decays back to two gluons, so in principle one can observe the $\ourpi$ as a narrow resonance in dijets. However,
due to the loop factor in the effective vertex, the $gg \to \ourpi$ cross section is strongly suppressed.  At the parton level, averaging over colors and
spins, we have
\beq\label{eq:ggtopi}
  \frac{1}{2^2} \frac{1}{8^2} \sum_{\rm color,spin} |{\cal M}_{gg\to \ourpi}|^2
  = \frac{15\alpha_s^2}{256\pi^2} \frac{\hat{s}^2}{f_\ourpi^2}  \>.
\eeq
Since we consider values of $m_{\ourpi}$ as low as $100\Gev$ we need to consider dijet resonance constraints from S$p\bar{p}$S. We integrate
(\ref{eq:ggtopi}) using CTEQ5L PDF's \cite{Lai:1999wy} to calculate the $\ourpi$ production cross section at a center of mass of $630\Gev$ and find
$\sigma(p\bar{p} \to \ourpi) \simeq 12\pb \>$ for $m_\ourpi = 100\Gev$ and $f_\ourpi = 43\Gev$ (i.e.~$m_\col=350\Gev$), which is below the bound given in
\cite{dijet-UA2} (for an earlier phenomenological study of colored resonances at S$p\bar{p}$S, see \cite{Berger:1984gk}). Similarly we obtain for Tevatron
Run-II $\sigma(p \bar{p} \to \ourpi) \simeq 3.8 \pb \>$ for $m_\ourpi = 250\Gev$ and $f_\ourpi = 110\Gev$ (i.e.~$m_\col = 830\Gev$). This is below the
dijet constraints of \cite{dijet-Teva} as can be seen also from figure \ref{fig:dijetexclusion}.

At Tevatron energies, one also needs to consider pair production of $\ourpi$, however note that even though we expect $2m_{\ourpi}<m_{\col}$, $\ourpi$
pair production is a $2$-$2$ process in contrast to resonant $\col$ production which both reduces the cross section and leads to a variation of
$\sqrt{\hat{s}}$ from event to event, thereby decreasing the significance of any excess in kinematic distributions. Therefore, we do not expect the search
strategy outlined in section \ref{sec:discov} to yield as high a significance for this process.

It is intriguing to contemplate how light a $\ourpi$-mass can be accommodated,
as $\ourpi$ couples only to gluons in the SM, thus most existing
experimental bounds are irrelevant. In this work we only consider $m_{\ourpi}
\gsim m_{Z}$ to avoid any constraints from corrections to the running of
$\alpha_{s}$.

\subsection{Other Sources of Potential Constraints}

Since neither $\col$ nor $\ourpi$ are electroweak charged, there are virtually
no constraints on our benchmark model from LEP direct searches or precision
electroweak data. Moreover, the fact that the $\col$-$q$-$\bar{q}$ coupling
arises via $\col$-gluon mixing makes the coloron coupling to quarks flavor
blind, therefore there are no constraints from flavor changing processes on our
benchmark model.

There are also no constraints from quark compositeness \cite{comp-bound}. This
is because compositeness bounds are sensitive to effective 4-fermion
operators arising from integrating out heavy particles, however the range of
coloron masses we consider is low enough for resonant production so the
compositeness bounds are replaced by the constraints from dijet resonance
searches, which are stronger.

One subtlety in our benchmark model is the existence of $\SU(3)_{\rm HC}$ baryons, the lightest of which is a color octet, just like the lightest QCD
baryons are arranged in an octet of flavor. Since hyper-baryon number $\U(1)_{\rm HB}$ is exact in our benchmark model, the lightest hyper-baryon (LHB) is
stable, while at collider time-scales the higher mass hyper-baryons decay promptly to the LHB. Once pair-produced, the LHB will hadronize with quarks or a
gluon to form a color-singlet. In fact, the LHB has the same quantum numbers as a (Dirac) gluino, and thus the limits on stable gluinos cited in
\cite{Mafi:1999dg} apply. Based on the close analogy between QCD and hyper-color, $m_{\rm LHB}\gsim m_{\col}$ holds, thus hyper-baryon pair production is
compatible with the bounds listed in \cite{Mafi:1999dg} for the range of parameters we consider in this work. We should note however, that it is
straightforward to include additional particle content with renormalizable couplings which causes the hyper-baryons to decay unobservably into SM singlets
plus jets at the Tevatron.

\section{Discovery Potential at the Tevatron}
\label{sec:discov}

Now that we have argued that our benchmark model is not ruled out by existing
experimental constraints, one may worry that it is simply not visible in any
channel, and hence not discoverable. This section will be aimed at showing that
this is not at all the case and that the Tevatron has a strong discovery
potential for our benchmark model within a broad range of parameters.

We will be concentrating in our search strategy on the production of the $\col$ particle, which is resonantly produced and dominantly decays to a pair of
$\ourpi$, which then decay to two pairs of gluons. Thus the background to consider is the 4-jet QCD background, which is both quite large in cross section
and has larger uncertainties compared to electroweak backgrounds. Fortunately there is one fact that favors signal over background for the range of $\col$
masses we are considering, namely that the signal is produced from a $q$-$\bar{q}$ initial state while the background is dominated by $g$-$g$ initiated
processes, and the valence quark PDF's do not fall as rapidly as the gluon PDF's at intermediate to high $x$ ($x\gsim0.2$). In any case, we will be
conservative in our analysis of signal significance estimates considering the uncertainties in background and we will look for evidence that manifests
itself as shape differences in kinematic distributions, in contrast to an excess in overall normalization. Anticipating an unknown $k$-factor in the
background we will use $2~\rm{fb}^{-1}$ of background in all our analyses even though we only use $1~\rm{fb}^{-1}$ of signal. We will use a simple
$\chi^{2}$ analysis to estimate the statistical significance of any excess, given by
\beq (\rm{stat.~sig.})^{2}=\sum_{bins}\left(\frac{n_{s}}{\sqrt{n_{b}}}\right)^{2} \eeq
where the sum extends over the entire distribution and bin size is chosen to be smaller than any kinematic features of the distribution but large enough
to contain many events.

Once again in the spirit of being conservative we limit our detailed studies to a region of parameter space where colorons are copiously produced, as our
main goal is to prove discoverability as well as to lay out a general search strategy. A detailed study conducted by experimentalists using more
sophisticated tools and statistical measures defining the limit of what is discoverable would be essential in producing concrete exclusion limits in the
event of non-discovery. As a case study we will use our benchmark model with two choices for the mass of the coloron, the first being relatively light
with $m_{\col}=350\Gev$ (where we take $m_ {\ourpi}=100\Gev$) and the second one being heavier with $m_{\col}=600\Gev$ (where we take $m_
{\ourpi}=180\Gev$). Colorons significantly lighter or heavier than these two cases are more challenging for discovery for reasons mentioned in our
conclusions.

In contrast to the background, where the jet energies in an event are usually hierarchical, we expect all four jets in signal events to have similar
energies. Therefore it appears plausible that a large cut on the $p_{T}$ of all four jets should reduce background more than signal, with the further
advantage that the perturbative QCD approximation employed in Monte Carlo simulations are more reliable for larger values of $p_{T}$. Moreover, for any
realistic study we have to take into account the triggers used in the Tevatron analysis in order to ensure that all events in our signal and background
samples are guaranteed to have been triggered on. To avoid issues with prescaled triggers we therefore will demand that all events used in our analysis
have at least one jet with $p_{T}\ge120\Gev$, thereby making certain that they would have passed the $100\Gev$ single jet trigger used in CDF
\cite{trigger}.

To simulate signal we use MadGraph version 4.2.3 \cite{MC} where we implement $\col$,$\ourpi$ and their relevant couplings to the SM using the provided
user-mode. We generate signal for an integrated luminosity of $1~\rm{fb}^{-1}$ using the process $p\bar{p}\to \ourpi\ourpi$. We then use the Pythia-PGS
interface \cite{MC}, where Pythia decays the $\ourpi$ into a pair of gluons, provides the parton shower and hadronization, and PGS is used for jet
reconstruction. We use the standard CDF parameter card supplied with the distribution, but use cone jets with $\Delta R=0.7$ in the reconstruction. For
background, we generate parton level events with MadEvent using the process $p\bar{p}\to jjjj$, and again use the Pythia-PGS interface with the same
parameters as for the signal.

\subsection{Lighter Coloron Case, $m_{\col}=350\Gev$}

For this choice of mass, we find the production cross section of the coloron to be $1.14\times10^{2}~\rm{pb}$, however only a fraction of signal passes
the leading jet $p_{T}$ cut of $120\Gev$ we are using to emulate the trigger, therefore we cannot afford to make too many other harsh cuts. We choose to
veto events which have less than 4 jets with $p_{T}$ greater than $40\Gev$. After these cuts, we find $\sigma_{s}=3.60~\rm{pb}$ while
$\sigma_{b}=65.8~\rm{pb}$.

\FIGURE[!t]{\includegraphics[width=5.0in]{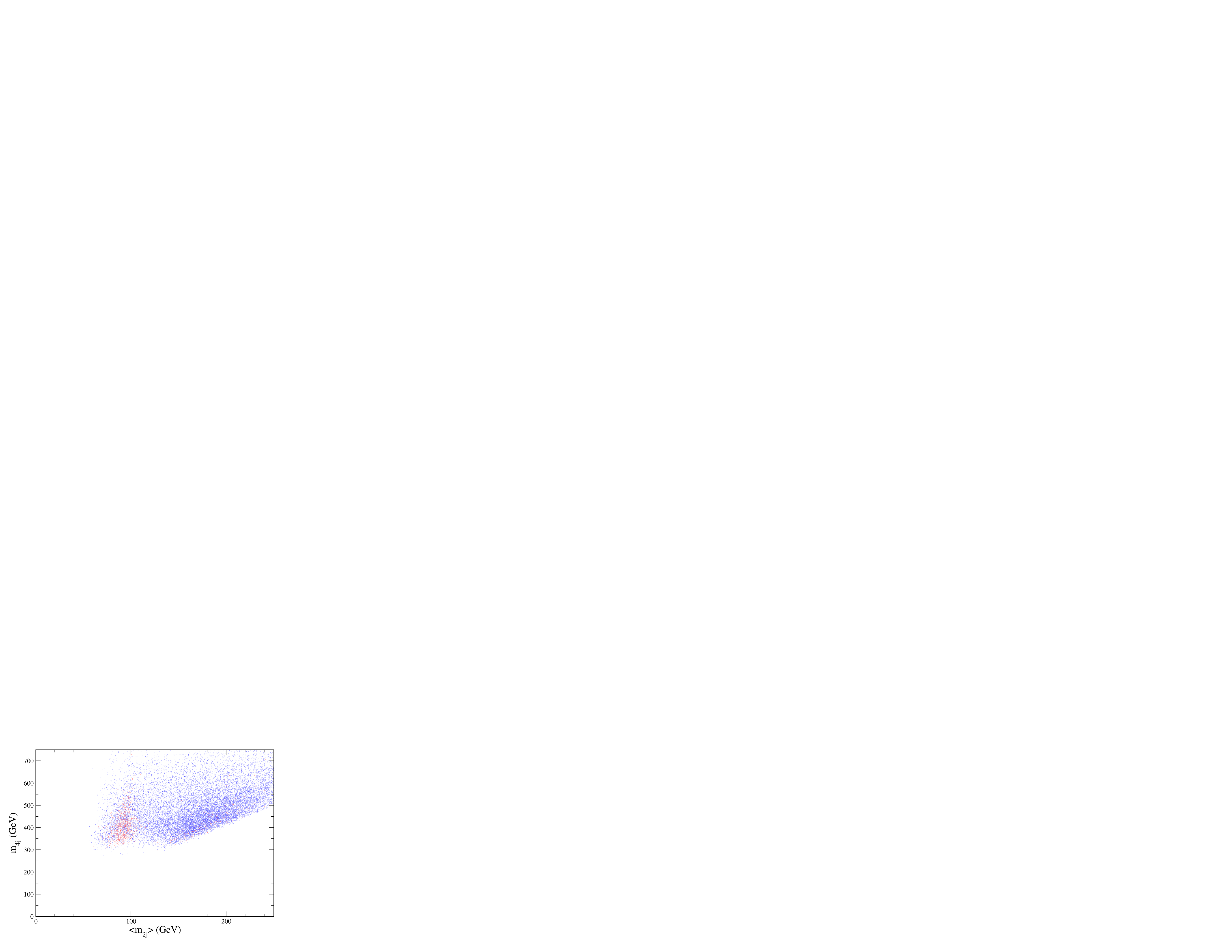} \caption{Dedicated coloron search in the benchmark model with $m_{\col}=350\Gev$ and $m_
{\ourpi}=100\Gev$ at Tevatron Run-II. We select events with at least one jet with $p_{T}>120\Gev$ and four jets with $p_{T}>40\Gev$ and we demand further
that the four jets can be paired such that the invariant mass of the pairs is within $25\Gev$ of each other. We then plot the average pair invariant mass
versus the 4j invariant mass. Each red dot represents a signal event which passed the cuts for $1~\rm{fb}^{-1}$ of integrated luminosity while each blue
dot represents a background event which passed the cuts for $2~\rm{fb}^{-1}$ of integrated luminosity. The red dots along the diagonal are mispaired
signal events, while most signal events are correctly paired and cluster near the true value of $(m_{\ourpi},m_ {\col})$.} \label{fig:scatterlight}}
To exploit the full kinematic information present in the signal we further pair the four leading jets into two pairs and veto all events where no possible
pairing yields two pairs with $m_{\rm inv}$ within $25\Gev$ of each other. (If there is more than one such possible pairing, we take the one that yields
the closest $m_{\rm inv}$ for the pairs.) This further reduces the signal cross section to $2.66~\rm{pb}$ and the background to $20.8~\rm{pb}$. We then
plot the average invariant mass of the two pairs against the invariant mass of the four leading jets. (A similar search strategy relying on the pairing of
four jets was used in \cite{Chivukula:1991zk} albeit without the additional advantage of the presence of a primary four jet resonance.) The results are
plotted in figure \ref{fig:scatterlight} where the shape difference between the signal and background is very clearly visible. Most signal points are
correctly paired and accumulate in a small region close to the actual masses of the $\col$ and $\ourpi$ while some signal events are mispaired and appear
scattered in a larger region along the diagonal where the background is most densely populated. We find the statistical significance of the excess to be
$32.3\,\sigma$. Even though we are aware that there are sources of systematic error that are not accounted for in our analysis, this result is strong
enough to indicate that such a search strategy will yield definitive results even when done with more sophisticated tools such as a fully realistic
detector simulation and taking into account shape dependent corrections or further subtleties involved in a real experimental analysis.

\FIGURE[!t]{\includegraphics[width=5.0in]{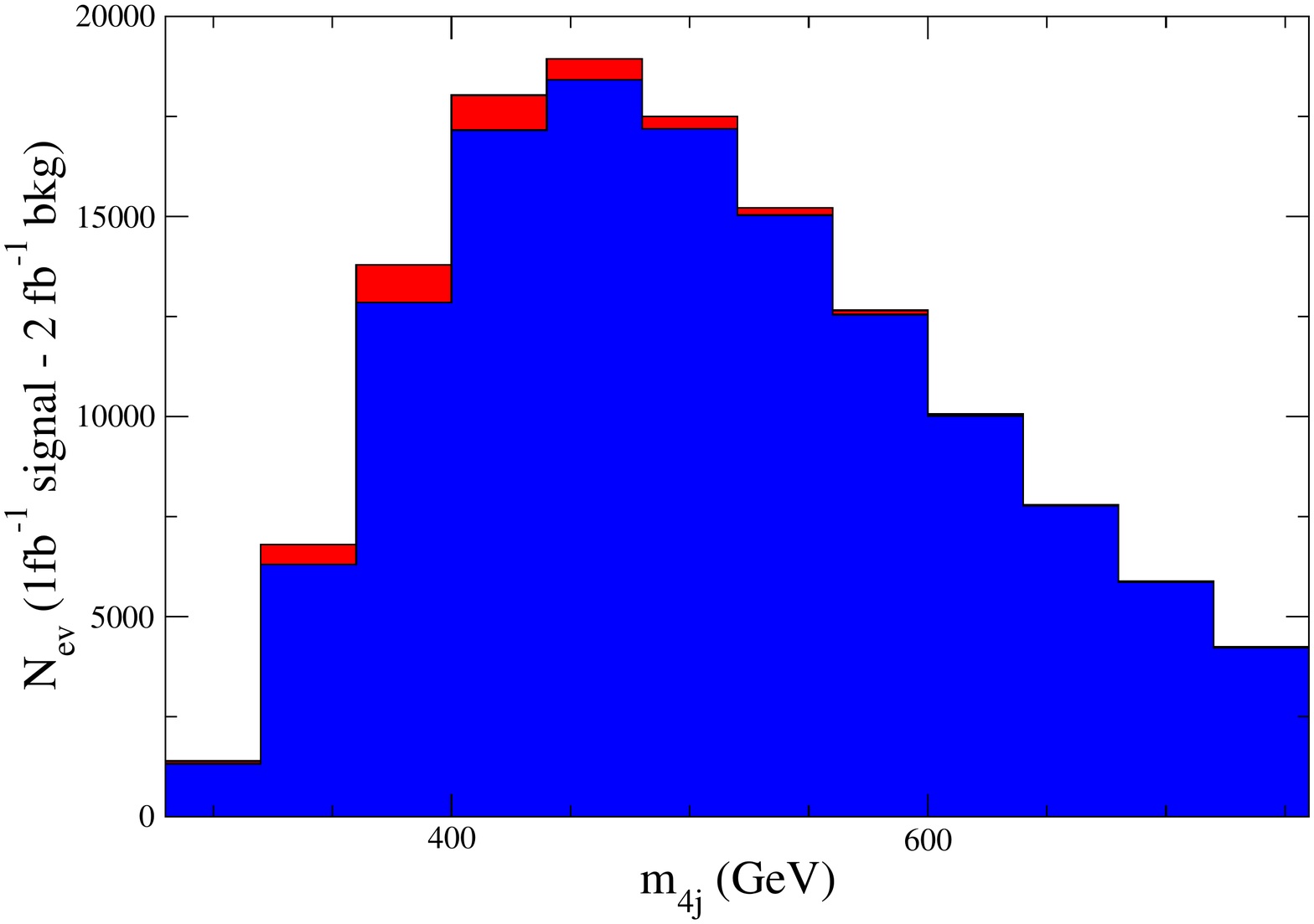} \caption{More general coloron resonance search in the 4j channel at Tevatron Run-II. In events
with at least one jet with $p_{T}>120\Gev$ and four jets with $p_{T}>40\Gev$ we plot the invariant mass of the four leading jets. Blue corresponds to
$2~\rm{fb}^{-1}$ of background while red corresponds to $1~\rm{fb}^{-1}$ of signal for $m_{\col}=350\Gev$.} \label{fig:inclusivelight}}
In fact, with such high signal significance it is interesting to attempt a less model dependent search that would have reduced sensitivity, which however
may be sensitive to models other than our benchmark, e.g. when the coloron decays to two particles of unequal mass. Therefore we try to be as inclusive as
possible and determine whether a search that was not optimized to look for secondary resonances would still discover the coloron. Using the same $p_{T}$
cuts as above but without pairing up the jets we simply construct the invariant mass of the leading four jets. The results are displayed in figure
\ref{fig:inclusivelight}. The significance of the excess in this distribution is $13.4\,\sigma$. In order to reduce any bias in the first few bins
introduced by analysis cuts we repeat the analysis where we disregard any discrepancy in the bins up to $m_{\rm inv}=400\Gev$ and still find a
significance of $8.3\,\sigma$.

Even though these results seem to suggest that an almost blind search could provide initial evidence for the existence of a colored resonance decaying to
a four jet final state, one needs to worry that corrections in the calculation of the background can give rise to a shape difference large enough to
nullify the significance of the excess in this more general search. This situation cannot be improved greatly in looking for the lighter coloron, since we
cannot make our $p_{T}$ cuts much harder without losing the signal. We will come back to this issue in the study of the heavier coloron however, and argue
that the prospects are much better in that case.

\subsection{Heavier Coloron Case, $m_{\col}=600\Gev$}

\FIGURE[!t]{\includegraphics[width=5.0in]{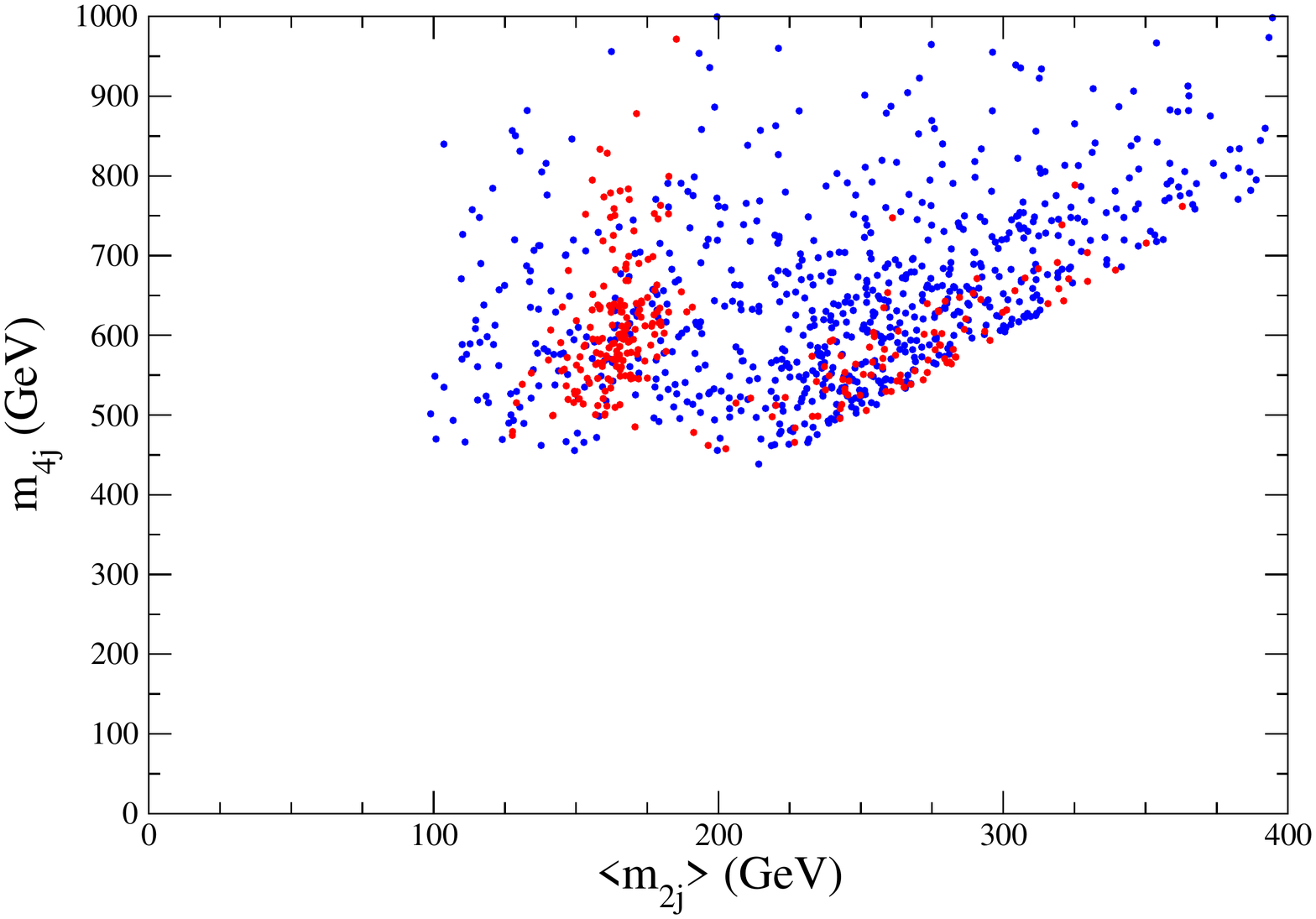} \caption{Dedicated coloron search in the benchmark model with $m_{\col}=600\Gev$ and $m_
{\ourpi}=180\Gev$ at Tevatron Run-II. We select events with at least one jet with $p_{T}>120\Gev$ and four jets with $p_{T}>90\Gev$ and we demand further
that the four jets can be paired such that the invariant mass of the pairs is within $25\Gev$ of each other. We then plot the average pair invariant mass
versus the 4j invariant mass. Each red dot represents a signal event which passed the cuts for $1~\rm{fb}^{-1}$ of integrated luminosity while each blue
dot represents a background event which passed the cuts for $2~\rm{fb}^{-1}$ of integrated luminosity.} \label{fig:scatterheavy}}
Having shown that even a coloron as light as $350\Gev$ can be discovered despite trigger inefficiencies for the signal as well as higher backgrounds, we
now study the case of a heavier coloron with $m_{\col}=600\Gev$ which has a production cross section of $10.0~\rm{pb}$ at Run-II. For this case, most
signal events automatically have a leading jet with $p_{T}\ge 120\Gev$ and we can afford to put a harder cut on the $p_{T}$ of the other jets. In fact we
will choose to accept events in our analysis which have at least four jets with $p_{T}\ge 90\Gev$. After these cuts, the signal and background cross
sections are $\sigma_{s}=0.36~\rm{pb}$ and $\sigma_{b}=0.99~\rm{pb}$. As before, we veto events in which the leading four jets cannot be paired in a way
to give two pairs with invariant masses within $25\Gev$ of each other, which further reduces the cross section after cuts to $\sigma_{s}=0.27~\rm{pb}$ and
$\sigma_{b}=0.38~\rm{pb}$. The results are displayed in figure \ref{fig:scatterheavy} where the significance of the excess is $17.2\,\sigma$.

\FIGURE[!t]{\includegraphics[width=5.0in]{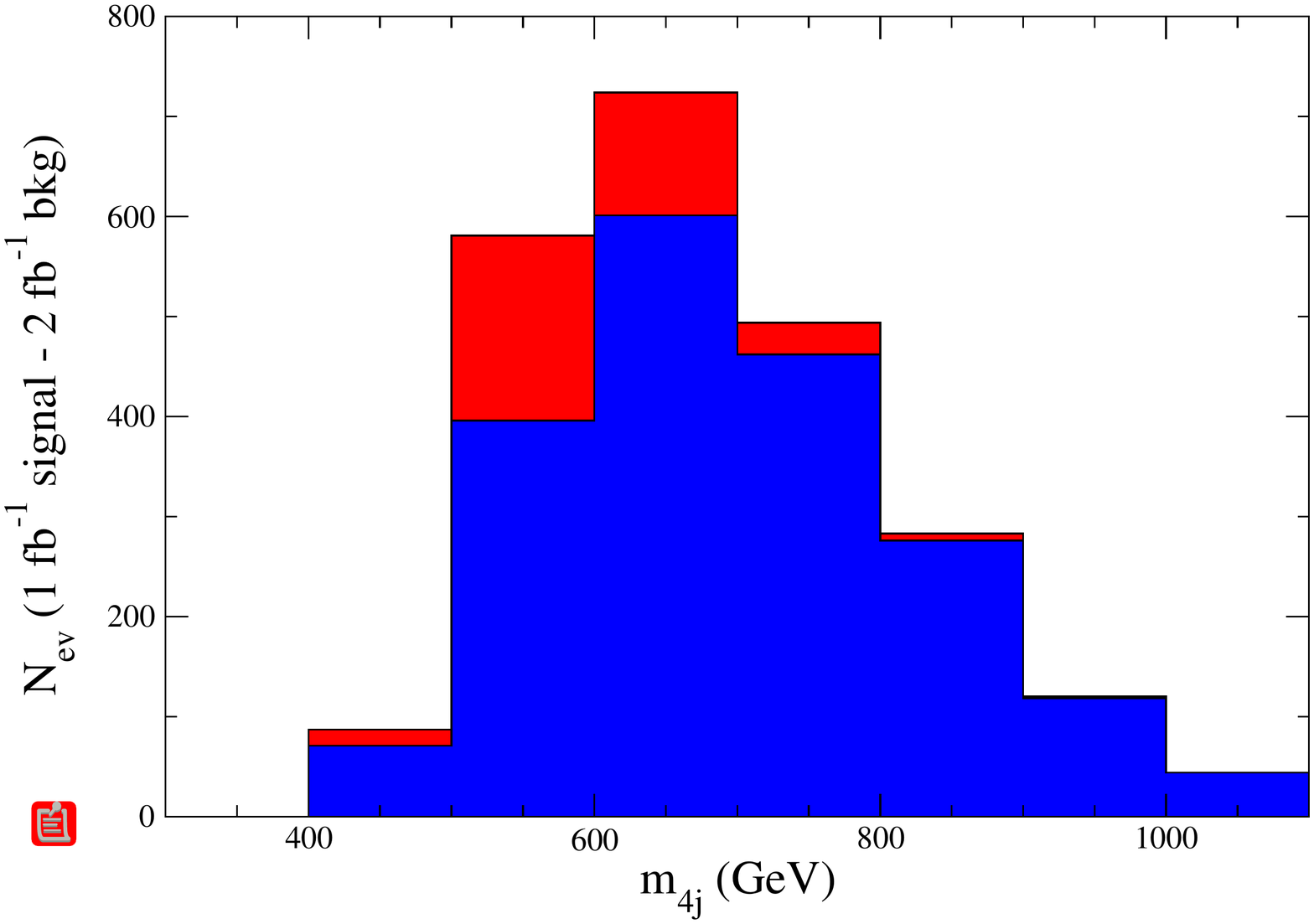} \caption{More general coloron resonance search in the 4j channel at Tevatron Run-II. In
events with at least one jet with $p_{T}>120\Gev$ and four jets with $p_{T}>90\Gev$ we plot the invariant mass of the four leading jets. Blue corresponds
to $2~\rm{fb}^{-1}$ of background while red corresponds to $1~\rm{fb}^{-1}$ of signal for $m_{\col}=600\Gev$.} \label{fig:inclusiveheavy}}
As before, we also perform a less model dependent search looking at the invariant mass of the four leading jets using the same cuts as above but without
demanding that they can be paired. The results are displayed in figure \ref{fig:inclusiveheavy} where the statistical significance of the excess is
$10.8\,\sigma$. Coming back to the issue of shape dependent corrections to the background we note that these are events where there are four very hard
jets which are maximally separated from each other, which is where we expect the perturbative expansion to be most reliable. Keeping in mind that we are
already using twice as much background as signal, it would require nearly a $100\%$ error on the shape of the background to eliminate the significance of
the signal excess in the case of the heavier coloron.

\section{Concluding Remarks}
\label{sec:conc}

We have emphasized in this paper how a variety of new physics scenarios can lead to the existence of a massive color octet vector meson, the coloron. We
have used an analogy to QCD to set up a benchmark model of a composite coloron and write a phenomenological Lagrangian for it, where our choices for the
values of the couplings are simply extrapolated from hadronic data. We have then shown that this benchmark model with new colored states at a few hundred
GeV is fully consistent with to-date experimental bounds and have outlined a promising search strategy at the Tevatron for discovering these states using
already existing data.

The range of coloron mass to which the Tevatron is sensitive can be understood as follows: If the coloron mass is too low (below about $300\Gev$), then
the signal events will not pass the single-jet trigger mentioned in section \ref {sec:discov} while prescaled triggers with lower thresholds would
severely reduce the signal significance. For coloron masses that are too large, the cross section drops below levels needed for discovery, for instance
above $850\Gev$ the cross section drops below $1~\rm{pb}$ which in our analysis would lead to a few tens of events after kinematic cuts. Exact limits on
the discovery reach will depend on a careful estimation of background, a realistic treatment of detector effects as well as a careful statistical
definition of what is discoverable, which should be part of a detailed study conducted by experimentalists. In this paper we have adopted a conservative
view of focusing attention on two choices for the coloron mass ($m_{\col}=350~\rm{GeV}$ and $m_{\col}=600~\rm{GeV}$ in the benchmark model) where we
demonstrated a strong potential for discovery even with the above mentioned uncertainties taken into account.

It is worth considering how the collider phenomenology is affected by the value of the ratio $m_{\ourpi}/m_{\col}$. This ratio is fixed in the benchmark
model, but in the spirit of \eq{eff-Lag} being an effective Lagrangian we can view $m_{\ourpi}$ as an independent parameter. Including input masses for
the hyper-quarks in \eq{UV-Lag} can increase $m_{\ourpi}/m_{\col}$ above $0.3$, its value in the benchmark model. Hyper-quark masses that are comparable
in magnitude to $\Lambda_{HC}$ will cause the $\col\rightarrow\ourpi\ourpi$ decay channel to become kinematically inaccessible, therefore this possibility
is ruled out by the dijet resonance constraints, on the other hand a modest increase in $m_{\ourpi}/m_{\col}$ will not significantly affect our analysis
methods. Smaller $\ourpi$ masses cannot be obtained from the microscopic theory \eq{UV-Lag} and even in the phenomenological model \eq{eff-Lag} the
benchmark value of $m_{\ourpi}$ is the technically natural one. If $m_{\ourpi}$ in \eq{eff-Lag} was tuned to a smaller value, one would expect the pairs
of jets resulting from the $\col$ decay to become more collinear, not only leading to a smaller efficiency in observing four separated hard jets and
making discovery more difficult, but in the extreme limit causing signal events to be reconstructed as having two back-to-back jets, thereby conflicting
with dijet resonance bounds.

We would like to elaborate on the point that the Tevatron is the ideal place to
look for a coloron, and that the LHC will have lessened sensitivity to a
signal of this kind. One reason is simply that the LHC, being a $p$-$p$
collider, will suffer from a suppression in the cross section of a resonance
produced from a $q$-$\bar{q}$ initial state especially for TeV scale
masses. The  lower mass ranges we considered in this paper would face severe
backgrounds at the LHC considering that the gluon PDF's would be sampled at much
lower $x$. Furthermore, the increased number of events per bunch crossing
due to higher luminosity as well as the higher trigger thresholds will make
event selection much more difficult for the LHC environment. On the other
hand, it would be interesting to assess the reach of the LHC for other states
appearing in the theory such as the hyper-baryon of our model, which can decay
to SM singlets plus jets, thereby leading to missing energy signatures, or even
be stable at collider time-scales in which case it would be challenging to
distinguish it from a stable gluino.

In this paper, the coupling of the coloron to ordinary quarks is inherited entirely via gluon-coloron mixing. This ensured compatibility of the new
physics with a variety of  precision tests, even for relatively low coloron mass. One can generalize this set-up to include  direct couplings of the
coloron (and/or secondary resonances) to quarks, either continuing in a flavor-blind manner \cite{coloron-2}, or flavor-dependently (in particular with
large top coupling) as motivated by some approaches to the Hierarchy Problem \cite{coloron-1}. When such direct couplings are  significant, the coloron
mass is forced above about a TeV in order to avoid constraints from some combination of low-energy precision tests,  dijet searches, and studies of top
production. That is, the coloron is effectively pushed outside the Tevatron window. However the direct couplings can then help compensate for some of the
difficulties associated with LHC discussed in the preceding paragraph, enabling discovery as dijet or top-antitop resonances, or in heavy flavor decays of
secondary resonances.  In this sense, the scenario discussed in the present paper and its Tevatron signatures is complementary to the case of significant
direct coloron-quark couplings and their LHC phenomenology.

Existing theoretical models occupy only a subset of possible phenomenological coloron parameters. Whether a coloron has a deep connection to the
(resolution of the) Hierarchy Problem or not, we cannot in generality predict the strength and flavor-dependence of coloron-quark couplings. While
different limits have now been studied theoretically, it may be that in an intermediate situation data from both the Tevatron and the LHC, in jets as well
as more distinctive channels such as heavy flavors or even missing energy, may prove essential in uncovering the physics of the coloron.

\begin{acknowledgments}
We would like to thank Johan Alwall for his infinite patience and help using MadEvent, Morris Swartz for invaluable discussions and advice, Sarah Eno,
Isaac Hall, Robert Harris, Kenichi Hatakeyama, Joey Huston, and Petar Maksimovic for assistance with experimental issues, Sekhar Chivukula, Bogdan
Dobrescu, David E.~Kaplan, Steve Mrenna, Matthew Schwartz, John Terning and Brock Tweedie for useful insights, Henry Tye for pointing out to us the
three-jet decay mode of the coloron, and Bruce Knuteson for helping us with issues concerning Tevatron global searches. The authors are supported by the
National Science Foundation grant NSF-PHY-0401513 and by the Johns Hopkins Theoretical Interdisciplinary Physics and Astrophysics Center. C.K.~and
T.O.~are further supported in part by DOE grant DE-FG02-03ER4127 and by the Alfred P.~Sloan Foundation. T.O.~is also supported by the Maryland Center for
Fundamental Physics.
\end{acknowledgments}


\end{document}